\documentclass[aps,eqsecnum,groupedaddress,fleqn]{revtex4}
\usepackage{graphicx}
\usepackage{amssymb}
\newcommand{\be}{\begin{equation}}
\newcommand{\ee}{\end{equation}}
\newcommand{\bea}{\begin{eqnarray}}
\newcommand{\eea}{\end{eqnarray}}
\newcommand{\bvec}[1]{\mbox{\boldmath $#1$}}
\newcommand{\eqref}[1]{(\ref{#1})}
\newcommand{\rot}[1]{ \bvec{\nabla} \times #1}

\begin{document}

\preprint{MPI-PhT/2003-06, HU-EP-03/05, KANAZAWA-03-04}

\title{Duality of gauge field singularities and the structure  
of the flux tube in 
Abelian-projected SU(2) gauge theory and the 
dual Abelian Higgs model}

\author{Y. Koma}
\email[]{ykoma@mppmu.mpg.de}
\affiliation{Max-Planck-Institut f\"ur Physik, 
F\"ohringer Ring 6, D-80805 M\"unchen,  Germany}

\author{M. Koma}
\email[]{mkoma@mppmu.mpg.de}
\affiliation{Max-Planck-Institut f\"ur Physik, 
F\"ohringer Ring 6, D-80805 M\"unchen,  Germany}

\author{E.-M. Ilgenfritz}
\email[]{ilgenfri@physik.hu-berlin.de}
\affiliation{Institut f\"ur Physik, Humboldt Universit\"at zu Berlin,
Newton-Str. 15, D-12489 Berlin,  Germany}

\author{T. Suzuki}
\email[]{suzuki@hep.s.kanazawa-u.ac.jp}
\affiliation{Institute for Theoretical Physics, Kanazawa University,\\
Kakuma-machi, Kanazawa, Ishikawa 920-1192, Japan}

\author{M.I. Polikarpov}
\email[]{polykarp@heron.itep.ru}
\affiliation{ITEP, B.Cheremushkinskaya 25, 
RU-117259 Moscow, Russia}

\date{\today}

\begin{abstract}
The structure of the flux-tube profile
in Abelian-projected (AP) SU(2) gauge theory 
in the maximally Abelian gauge is studied.
The connection between the AP flux tube and
the classical flux-tube solution 
of the U(1) dual Abelian Higgs (DAH) model
is clarified in terms of the path-integral 
duality transformation.
This connection suggests that the electric photon 
and the magnetic 
monopole parts of the Abelian Wilson loop
can act as separate sources creating the Coulombic 
and the solenoidal electric field inside a flux tube.
The conjecture is confirmed by a lattice simulation
which shows that the AP flux tube is composed of 
these two contributions.
\end{abstract}

% \pacs{12.38.Aw, 12.38.Gc, 11.15.Ha}
% \keywords{bbb}
\maketitle

\baselineskip=0.5cm
\parskip=0.4cm

%%%%%%%%%%%%%%%%%%%%%%%%%%%%%
\section{Introduction}

\par
When the QCD vacuum is viewed 
as a {\em dual} superconductor~\cite{tHooft:1975pu,Mandelstam:1974pi},
the quark confinement mechanism 
can be immediately understood:
the (color-) electric flux associated with a quark-antiquark 
($q$-$\bar{q}$) system is squeezed into an 
almost-one-dimensional flux tube by the {\em dual} 
Meissner effect caused by magnetic monopole condensation.
This picture leads to a linear confinement potential and
is a dual analogue of the magnetic Abrikosov vortex in an ordinary 
superconductor~\cite{Abrikosov:1957sx,Nielsen:1973ve,Nambu:1974zg}.
It is natural to expect
that it can be quantitatively formulated 
by a dual version of an Abelian Higgs model,
the dual Abelian Higgs (DAH) model.
The Lagrangian
--- besides the kinetic terms of each field and a minimal 
coupling between the two fields ---
should contain a monopole self-interaction term 
that allows for a broken phase of dual gauge symmetry.
The DAH model indeed has an electric flux-tube solution 
of the static $q$-$\bar{q}$ system~\cite{Nambu:1974zg}.

\par
A linear potential emerging from a flux tube is quite welcome
to give an interpretation 
for the area law behavior of the Wilson loop 
observed in lattice QCD simulations~\cite{Creutz:1984mg}.
It would explain 
the Regge trajectory pattern or other string-like 
properties of hadrons~\cite{Sailer:1991jv}.
Then the problem arises how to derive the
dual superconductor scenario from QCD, that is,
how to formally derive the DAH model from QCD.
One also would like 
to observe certain characteristic
features of the dual superconductor,  
such as the formation of flux tubes, through a 
Monte-Carlo simulation of lattice QCD
directly.

\par
As for the formal derivation, it is known that if magnetic monopoles 
are introduced as the consequence of Abelian projection
{\it a la} 't~Hooft~\cite{tHooft:1981ht}
and if the diagonal components of gluons play a dominant role 
(compared to the off-diagonal ones) in the 
long distance behavior of QCD (Abelian dominance),
a condensed phase of monopoles is realized beyond a certain critical 
scale~\cite{Ezawa:1982bf,Ezawa:1982ey}.
Remarkably, lattice QCD simulations with non-Abelian configurations
undergoing  't~Hooft's Abelian-projection
(typically in the maximally Abelian gauge, MAG)  
support this scenario numerically.
For instance, the string tension measured by the
``Abelian Wilson loop'' 
constructed from the Abelian link variables
(the ``Abelian string tension''), 
is almost saturating the
non-Abelian string tension~\cite{Suzuki:1990gp}.
In this context, 
applying the Zwanziger formalism~\cite{Zwanziger:1971hk}, 
one can introduce the dual gauge field which 
is minimally coupled 
to monopoles. Summing over monopole current 
trajectories~\cite{Bardakci:1978ph,Stone:1978mx},  
one can also introduce a monopole field.
This formulation finally leads to the DAH 
model~\cite{Suzuki:1988yq,Maedan:1988yi,%
Suganuma:1995ps,Sasaki:1995sa}.
However, it is difficult to determine the effective 
couplings of the DAH model through this 
analytical derivation, because 
one cannot treat the monopole current system
quantitatively.
In order to accomplish this, one would 
need numerical investigations of 
monopole dynamics on the lattice, 
for instance by means of the inverse Monte-Carlo 
method~\cite{Shiba:1995pu,Shiba:1995db,Kato:1998ur,Chernodub:2000ax}.
This might require more complicated ans\"atze for matching
the monopole actions~\cite{Suzuki:2001tp}.

\par
Just in order to seek
flux-tube configurations in the non-Abelian gauge theory,
the profiles of the electric field and the monopole current 
distribution induced by an Abelian Wilson
loop  have been studied within the Abelian-projection 
scheme~\cite{Singh:1993jj,Matsubara:1994nq,Cea:1995zt,Bali:1998gz}.
It has been found that the shapes are similar to
those of the flux-tube solution in the DAH model.  From 
now we call the former one ``Abelian-projected (AP) flux tube''
and the latter one ``DAH flux tube''.
We remark that the connection between the 
AP flux tube and the DAH flux tube is not 
on equal footing because the former contains the quantum effects
at work in non-Abelian lattice gauge simulations
on the original lattice, 
while the latter is a classical solution obtained
by solving the field equations
with dual variables.  
Having in mind this conceptual difference,
it is still worth to
determine the effective couplings of the DAH model, which 
could not be fixed through a formal derivation,
through the comparison between the two flux tubes.
This is interesting because, once the DAH parameters are fixed, 
one can use the DAH model for further analyses:
for discussing hadronic 
objects~\cite{Kamizawa:1993hb,Koma:1999sm,Koma:2000hw},
for investigating the dynamics of the flux tube by deriving 
an effective string action from the DAH 
model~\cite{Akhmedov:1996mw,Antonov:1998xt,Antonov:1998wi,%
Chernodub:1998ie,Koma:2001pz,Koma:2002rw}, etc.

\par
Up to now, the quantitative status of the comparison
between the AP and the DAH flux tubes has not been conclusive,
although this has been attempted
several times~\cite{Singh:1993jj,Matsubara:1994nq,Cea:1995zt,%
Bali:1998gz,Gubarev:1999yp}.
In order to find DAH parameters which 
possess physical meaning in this context, 
at first it is important to understand to 
what extent the AP flux tube can be 
really related to the DAH flux tube,
first of all since they are defined 
in terms of different (original and dual) variables.
This should become clear 
once the duality transformation is carried out in detail.
Second, also a more systematic study of the AP flux-tube 
profile is required to have well-controlled lattice data;
one needs to check the Gribov copy effect 
hidden in the process of MAG fixing, 
has to examine to what extent the scaling property is fulfilled, 
should investigate the $q$-$\bar{q}$ distance dependence of the flux 
tube shape etc. on a sufficiently large lattice volume.

\par
In this paper, we aim to address 
only the first part, the qualitative 
and detailed relation between the AP and the DAH flux tubes.
Here we do not attempt to fix the DAH model parameters.
What we plan to do here is to 
show that the AP flux tube has the 
composed internal structure as the DAH flux tube has, 
going through the path-integral 
duality transformation of the AP gauge theory.
In fact, in the DAH model, as we explain later in detail, 
the appearance of the electric flux tube is due to 
the superposition of two well distinguished components,
a Coulombic electric field, directly induced by  
the electric charges, and a solenoidal electric 
field induced by a monopole supercurrent. 
They are responsible for the Coulombic and the linearly rising part,
respectively, of the inter-quark 
potential in the DAH model.
If the electric flux profile can be uniquely
decomposed in the case of the AP flux tube as well,
analogously to the DAH flux tube, this will be an additional
argument in favor  of the DAH model description,
which will be important
for further quantitative discussions. 

\par
The guiding idea to discover this kind of structure also in the AP
flux tube comes from the measurement of the $q$-$\bar{q}$ potential
in terms of the Abelian Wilson loop.
The investigation of the Abelian Wilson loop
using the decomposition into an electric photon part (``photon Wilson loop'')
and a magnetic monopole part (``monopole Wilson loop'')
shows that also the Abelian potential consists of a Coulombic and 
a linear potential~\cite{Stack:1994wm,Shiba:1994ab,Bali:1996dm}.
We notice that this structure is
quite similar to that of the $q$-$\bar{q}$ potential in the DAH model.

\par
The paper is organized as follows.
In section~\ref{sec:2} we shall discuss the 
theoretical connection of the photon and 
the monopole Wilson loops with the composed 
structure of the DAH flux tube. 
We do this by closely looking at 
the path-integral duality transformation 
of the AP gauge theory.
Motivated by lattice results on the effective 
monopole action
we adopt, as our starting point,
a Villain type compact QED
as {\it the} approximate action of the 
{\it effective, AP gauge theory}.
In section~\ref{sec:3} 
we present the numerical results, the flux profile
induced by the photon and monopole Wilson loops,
measured within SU(2) lattice gauge theory in the MAG.
We come to the conclusion that the AP flux tube
is composed out of Coulomb and solenoidal parts, which 
add up to the full electric flux tube, in the same manner
as the DAH flux tube.
Section~\ref{sec:4} is the summary.

\par
The due improvement in the systematic study of the AP flux tube
including all details of the quantitative analysis of our lattice data, 
along the guiding lines formulated in the present paper,
is the subject of our follow-up paper~\cite{koma-next}.

%%%%%%%%%%%%%%%%%%%%%%%%%%%%%%%%
\section{The composed structure of the flux tube in the DAH model}
\label{sec:2}

\par
In this section, based on a path-integral analysis,
we discuss a possible theoretical relation between 
the electric-photon and magnetic-monopole parts
of the Abelian Wilson loop in the AP-SU(2) 
lattice gauge theory and the composed internal structure of 
the flux-tube solution in the U(1) DAH model.

\par  From lattice 
studies of the effective monopole action in the 
MAG~\cite{Shiba:1995pu,Shiba:1995db,Kato:1998ur,Chernodub:2000ax}, 
it is numerically suggested that, at some infrared scale,
the partition function of the AP-SU(2) theory 
is represented by the Villain type 
modification of compact QED.
Thus, we regard it as the effective AP gauge theory
and start from the partition function
\be
{\cal Z} = 
\int_{-\pi}^{\pi}
{\cal D} \theta
\sum_{n^{(m)} \in Z\!\!\! Z}
\exp 
\left [-  \frac{1}{2}  (F, \Delta D \; F) + i(\theta,j) 
\right ]\; .
\label{eqn:villain-type}
\ee
$F(C_{2})$ is the field strength
\be
F = d\theta -2 \pi n^{(m)} \; ,
\label{eq:def_F}
\ee
which is composed of compact link variables,
$\theta (C_{1}) \in [-\pi,\pi)$,
and  magnetic Dirac strings, 
$n^{(m)}  (C_{2}) \in Z\!\!\!Z$~\cite{Chernodub:1997ay}.
$\theta$ corresponds to the Abelian gauge field,
which interacts with 
an external electric current $j (C_{1}) \in Z\!\!\! Z$.
The operator $D$ is a general differential operator
and $\Delta$ is the Laplacian on the lattice.
In the infrared limit, it is 
numerically shown that 
the operator $D$
is well-described by the following form:
$D = \beta_{e} \Delta^{-1}+ \alpha + \gamma \Delta$,
where $\beta_{e}$, $\alpha$ and $\gamma$ 
are renormalized coupling constants of the 
monopole action which satisfy the relation 
$\beta_{e} \gg \alpha$, $\gamma$~\cite{Suzuki:2002sr}.
The (inverse) effective gauge coupling is
$\beta_{e} \equiv 4/e^{2}$. 
Since the magnetic Dirac strings $n^{(m)}$
are bordered by magnetic monopole currents 
$k (C_{3})$ as $d n^{(m)}  = -k$ (hence $dk=0$),
the Abelian Bianchi identity is now violated as
$d F = -2 \pi \; d n^{(m)}  = 2\pi k$.

\par 
For a conserved electric current, $\delta j=0$, we call
\be
W_{A}[j] \equiv \exp [i(\theta,j)]
\label{eqn:abelianloop}
\ee
the Abelian Wilson loop.
Its electric-photon ($W_{\mathit{Ph}}$)
and the magnetic-monopole 
($W_{\mathit{Mo}}$) parts are specified as follows.
Applying the Hodge decomposition to $\theta$,
we have the relation
\bea
\theta 
&=& \Delta^{-1}\Delta \theta \nonumber\\*
&=&  
\Delta^{-1}( \delta d + d\delta) \theta \nonumber\\*
&=&
  \Delta^{-1} \delta F
+ 2 \pi \Delta^{-1} \delta n^{(m)}
+ \Delta^{-1} d\delta \theta \nonumber\\*
&=&
 \Delta^{-1} \delta F
+ 2 \pi \Delta^{-1} \delta p
+ 2 \pi q
+ \Delta^{-1} d\delta  (\theta -2 \pi q).
\label{eqn:link-decompose}
\eea
In the last line, we have used the relation $n^{(m)}  = p + d q$,
where $p (C_{2}),\; q(C_{1}) \in Z\!\!\!Z$.
This means that 
an arbitrary shape of the open 
magnetic Dirac string $n^{(m)}$ is 
in general described by the sum of 
a fixed open string  $p$ with $dp=-k$
and the  closed strings $d q$ with $d^{2}q =0$.
Since all possible closed string fluctuations are
summed over, one can choose an arbitrary 
open string $p$.
Inserting Eq.~\eqref{eqn:link-decompose} into 
Eq.~\eqref{eqn:abelianloop},
the Abelian Wilson loop can be written as
\be
W_{A} [j] 
= \exp [i (\Delta^{-1} \delta F, j )]
\cdot 
\exp [i (2 \pi \Delta^{-1} \delta p, j )]
=
W_{\mathit{Ph}}[j]
\cdot 
W_{\mathit{Mo}}[j],
\label{eq:wilson-decomposition}
\ee
where the third and fourth terms
of Eq.~\eqref{eqn:link-decompose} do not contribute 
to this decomposition because of the relations 
$\exp[ 2 \pi i (q,j)]=1$ and $\delta j=0$.

\par
Let us proceed 
with the path integration of the 
partition function~\eqref{eqn:villain-type}
{\em keeping track} of the two parts of the Wilson loop, 
$W_{\mathit{Ph}}[j]$ and $W_{\mathit{Mo}}[j]$.
For simplicity and for picking up
the essence of the following discussions,
we restrict the differential operator 
in Eq.~\eqref{eqn:villain-type} to the leading term, 
$D = \beta_{e} \Delta^{-1}$.
The path integral duality transformation of such a model itself
has been discussed in many places since the 
works~\cite{Banks:1977cc,Peskin:1978kp}.

\par
We first rewrite the summation over Dirac strings 
as the independent summation over monopole 
currents $k$ (with constraint $d k=0$)
and $q$ as
\be
\sum_{n^{(m)}   \in Z\!\!\!Z}
=
\sum_{k \in Z\!\!\!Z, \; d k=0}
\;
\sum_{q  \in Z\!\!\!Z}\; .
\label{eq:sum}
\ee
Then, the integration with respect to $\theta$ is replaced by 
\bea
\int_{-\pi}^{\pi}{\cal D} \theta \sum_{q \in Z\!\!\!Z}
=
\int_{\infty}^{\infty} {\cal D}\theta^{ph} \; ,
\eea
where $\theta^{ph}= \Delta^{-1} \delta F$
represent noncompact link variables. 
Acting with an exterior derivative on 
$\theta^{mo}= 2 \pi \Delta^{-1}\delta p$, 
one finds $d \theta^{mo} = 2 \pi (n^{(m)}  + C^{(m)}  - d q)$ 
with $C^{(m)}  = \Delta^{-1}\delta k$.
Thus, the partition function is written as
noncompact QED
with summation over closed monopole currents, 
\be
{\cal Z}=
\int_{-\infty}^{\infty}
{\cal D} \theta^{ph} \sum_{k \in Z\!\!\! Z, \; dk =0}
\exp \left [-  \frac{\beta_{e}}{2} 
(d\theta^{ph}+ 2\pi C^{(m)} )^{2}
+i(\theta^{ph}+ \theta^{mo},j) 
\right ]\; .
\ee
In this expression one still realizes the 
violation of Abelian Bianchi identity in the form 
$d F = 2 \pi d C^{(m)} =2 \pi k$
due to  $d C^{(m)}  = k$.
Using the relation $(d\theta^{ph}, C^{(m)})
=(\theta^{ph},\delta  C^{(m)})=0$ (since $\delta  C^{(m)} =0$)
one can write $(F)^{2} = (d \theta^{ph})^{2} + 4 \pi^{2} (C^{(m)} )^{2}$.
Taking into account the gauge fixing condition
$\delta \theta^{ph}=0$,
one can integrate over $\theta^{ph}$. 
This yields a direct interaction term between
electric currents $j$ via the Coulomb 
propagator $\Delta^{-1}$.
Thus we have 
\bea
{\cal Z}
&=&
\sum_{k \in Z\!\!\! Z, \; d k =0} 
\!
\exp \left [ - 
\frac{1}{2\beta_{e}} (j,\Delta^{-1} j ) 
- 2 \pi^{2}\beta_{e} (C^{(m)} )^{2} + i(\theta^{mo},j)
\right ]\; .
\label{eq:splitting_interactions}
\eea
Defining $C^{(e)}  ({}^{*\!}C_{2}) \equiv \Delta^{-1} \delta * j $ 
in analogy to $C^{(m)} = \Delta^{-1}\delta k$, 
the first term of the action can also be 
written in the form
\be
\frac{1}{2\beta_{e}} (j,\Delta^{-1} j ) 
=\frac{1}{2\beta_{e}} ( *j,\Delta^{-1}  *j ) 
= \frac{1}{2\beta_{e}}  (C^{(e)})^{2} 
= 2 \pi^{2}\beta_{m}  (C^{(e)})^{2} \; ,
\label{eq:C_e_definition}
\ee
where we have introduced the (inverse) dual gauge coupling 
$\beta_{m}=1/g^{2}$, which should satisfy 
$4\pi^{2} \beta_{e}\beta_{m}=1$ ({\it i.e.} Dirac's condition  $eg=4\pi$).
Similarly, the square of  $C^{(m)}$ can be rewritten
\be
2 \pi^{2}\beta_{e}  (C^{(m)})^{2}  
= 
\frac{1}{2 \beta_{m}} ( k,\Delta^{-1} k) 
=
\frac{1}{2 \beta_{m}} ( *k,\Delta^{-1} * k) 
\; .
\label{eq:C_m_redefinition}
\ee
The exponential of this expression can further be 
understood as 
resulting from functional integration over the magnetic part of a 
noncompact {\em dual} gauge field 
$\tilde{\theta}^{mo}  ( {}^{*\!}C_{1})$,
minimally coupled to the magnetic monopole current,
\be
\exp \left [ - \frac{1}{2 \beta_{m}} ( *k,\Delta^{-1} *k) \right]
=
\int_{-\infty}^{\infty}
{\cal D}\tilde{\theta}^{mo}
\exp \left [ - 
\frac{\beta_{m}}{2} (d \tilde{\theta}^{mo})^{2} 
+
i (\tilde{\theta}^{mo}, *k ) 
\right ]\; .
\label{eq:path_integral_for_C_m}
\ee
We have attached the superscript `` ${}^{mo}$ ''
in order to distinguish it from 
the photon part of the dual gauge field, 
$\tilde{\theta}^{ph}  ( {}^{*\!}C_{1})$, 
which is defined in analogy to $\theta^{mo}$ as 
\be
\tilde{\theta}^{ph} =  2 \pi \Delta^{-1} \delta  n^{(e)}.
\label{eqn:dual-photon}
\ee
Here $n^{(e)} ({}^{*\!}C_{2}) \in Z\!\!\! Z$ denotes 
electric Dirac strings, satisfying $d n^{(e)}= - * j$,
which necessarily accompanies the presence of external 
electric charges.
Formally, $\tilde{\theta}^{ph}$ enters our consideration
when we re-express the monopole Wilson loop, using the relation
\be
(\theta^{mo} ,j ) - (\tilde{\theta}^{ph}, *k) 
= - 2 \pi ( p , * n^{(e)}) = 2 \pi N 
\qquad (N \in Z\!\!\! Z) \; .
\label{eq:intersection}
\ee
This means that the direct coupling of $j$ to 
$\theta^{mo}$ can be 
set equal to that of $k$ to $\tilde{\theta}^{ph}$, because
of $\exp[i(2\pi N)]=1$.
Thus, the partition function is found to be
\bea
{\cal Z} = 
\int_{-\infty}^{\infty}
{\cal D} \tilde{\theta}^{mo} 
\sum_{k \in Z\!\!\! Z, \; d k =0}
\exp \Biggl [ - \frac{\beta_{m}}{2} 
(d \tilde{\theta}^{mo} + 2 \pi C^{(e)} )^{2}
+ i (\tilde{\theta}^{mo} +\tilde{\theta}^{ph}, * k)  
\Biggl ] \; .
\label{eq:Z_dual_B_only}
\eea
The action is invariant under the transformation
$\tilde{\theta}^{mo} \mapsto \tilde{\theta}^{mo} 
+  d \tilde{f}$. 
This is nothing but the realization of the dual gauge 
symmetry, due to the 
conservation of magnetic monopole currents, $d k =0$.
In this action, the electric currents
are now {\em implicitly} defined via the violation of the
{\it dual} Abelian Bianchi identity written down for 
the dual field strength 
\be
\tilde{F} = d \tilde{\theta}^{mo} + 2 \pi C^{(e)}
\label{eq:dual_F}
\ee
as $d \tilde{F} = 2\pi d C^{(e)} = 2 \pi * j$, 
where $d C^{(e)}= * j$.

\par
The summation over monopole currents 
is the most difficult part of the evaluation.
In principle, one needs to know the monopole 
dynamics, for instance, 
such as monopole current distribution in the vacuum
and self interactions, etc.
The numerical investigations 
of the effective monopole actions
based on lattice Monte-Carlo simulation in the MAG
provide such information, which has suggested
the approximate form of the AP action 
given in Eq.~\eqref{eqn:villain-type}.
Here, we are not going to deal with these 
complication, since 
the kinetic structure of the dual gauge field,
being composed of a regular $\tilde{\theta}^{mo}$ and 
a singular $\tilde{\theta}^{ph}$ parts,
is not affected by the summation over monopoles.
We then assume that the monopole current system 
is described by the grand canonical ensemble 
of closed loops, interacting via the dual gauge field.
Then the complex-valued 
scalar monopole field $\chi$,
which minimally couples to the dual gauge field,
is introduced~\cite{Bardakci:1978ph,Stone:1978mx} 
instead of monopole currents as
\bea
\sum_{k \in Z\!\!\! Z, \; d k =0}
\exp \Biggl [ i (\tilde{\theta}^{mo} +\tilde{\theta}^{ph}, * k)  
\Biggl ] 
\to
\int {\cal D} \chi  {\cal D} \chi^{*}
\exp 
\Biggl [ -  \left \{ | (d + i (\tilde{\theta}^{mo} 
+\tilde{\theta}^{ph}) ) \chi |^{2}  
+ \lambda (|\chi|^{2}-v^{2})^{2} \right \}
\Biggl ] ,
\label{eqn:currentsum}
\eea
where the $\lambda |\chi|^{4}$  ($\lambda >0$) term 
plays the role to keep the density of loops being finite 
(it produces a short distance repulsion 
between the loop segments)
and $v$ denotes monopole condensate which 
describes the typical scale of the system.
In this way, we arrive at the DAH model,
\be
S_{\rm DAH} = \frac{\beta_{m}}{2} 
( \tilde{F})^{2}
+
| (d + i (\tilde{\theta}^{mo} 
+\tilde{\theta}^{ph}) ) \chi |^{2}  
+ \lambda (|\chi|^{2}-v^{2})^{2}.
\ee
Although we cannot argue the precise values of 
the effective couplings in this formal derivation,
we can restrict ourselves to the 
range of parameters able
to describe the condensed phase of monopoles,
according to the lattice results~\cite{Shiba:1995db}.

\par
Due to the singular structure 
of $\tilde{\theta}^{ph}$ associated with $n^{(e)}$
(see, Eq.~\eqref{eqn:dual-photon}),
the DAH model {\em has} the open flux-tube solution,
obtained by solving the field equations,
\bea
&&
\beta_{m} \partial_{\mu} 
(\partial_{\mu} \tilde{\theta}_{\nu}^{mo}
-\partial_{\nu} \tilde{\theta}_{\mu}^{mo} )
= 2 (\tilde{\theta}_{\nu}^{mo} + \tilde{\theta}_{\nu}^{ph})  \phi^{2} 
= k_{\nu}
\; ,
\label{eq:feq-1-m}\\
&&
\partial_{\mu}\partial_{\mu} \phi
+ (\tilde{\theta}_{\mu}^{mo} 
+\tilde{\theta}_{\mu}^{ph})^{2}
\phi
=
 2 \lambda \phi ( \phi^{2} - v^2) \; .
\label{eq:feq-2-m}
\eea
Here, we have inserted
the polar decomposition of the monopole field
$\chi=\phi \exp (i \eta)$ ($\phi,\eta \in \Re$),
and the phase $\eta$ has been absorbed 
into the definition of $\tilde{\theta}_{\mu}^{mo}$.
The boundary conditions of the dual gauge field
and monopole field are determined so as
to make the energy of the system finite:
just on the electric Dirac string $n^{(e)}$,
$\tilde{\theta}^{mo}_{\mu} =0$ and $\phi =0$ 
whereas at large distance from the string,
$\tilde{\theta}^{mo}_{\mu} = - \tilde{\theta}^{ph}_{\mu}$
and $\phi=v$.
After solving the field equations (in general, numerically),
we can compute the profile of the electric field 
as the spatial part of the field strength
$\tilde{F}$ in Eq.~\eqref{eq:dual_F},
\bea
\bvec{E}
&=&
\rot{\tilde{\bvec{\theta}}}^{mo} + 2 \pi \bvec{C}^{(e)}
\equiv 
\bvec{E}^{mo} + \bvec{E}^{ph},
\label{eq:electricprofile}
\eea
and the magnetic current as the 
spatial part of the monopole current, 
\bea
\bvec{k}
&=&
2 (\tilde{\bvec{\theta}}^{mo} 
+\tilde{\bvec{\theta}}^{ph} ) \phi^{2}, 
\label{eq:magneticprofile}
\eea
respectively.
Concrete forms of the field equations and 
the boundary conditions of fields for the
straight $q$-$\bar{q}$ system are given in the 
Appendix~\ref{sec:DAHfluxtube}.

\begin{figure}[!t]
\centering
\includegraphics[width=12cm]{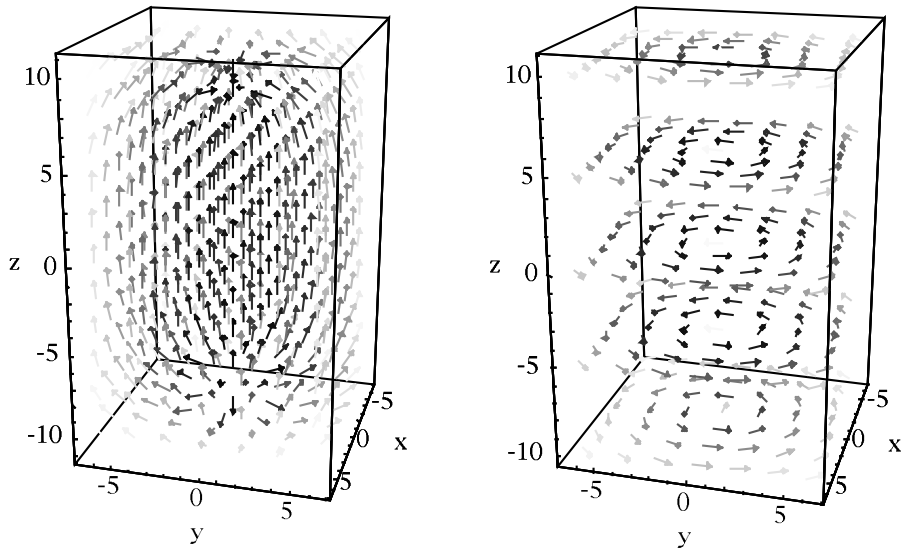}
\caption{Profiles of electric field 
$\bvec{E}\cdot a^{2}$ (left) and monopole 
current $\bvec{k}\cdot a^{3}$ (right) in the DAH model,
where $a$ denotes a certain length scale.} 
\label{fig:dgl-profile}
\end{figure}

\begin{figure}[!t]
\centering
\includegraphics[width=15cm]{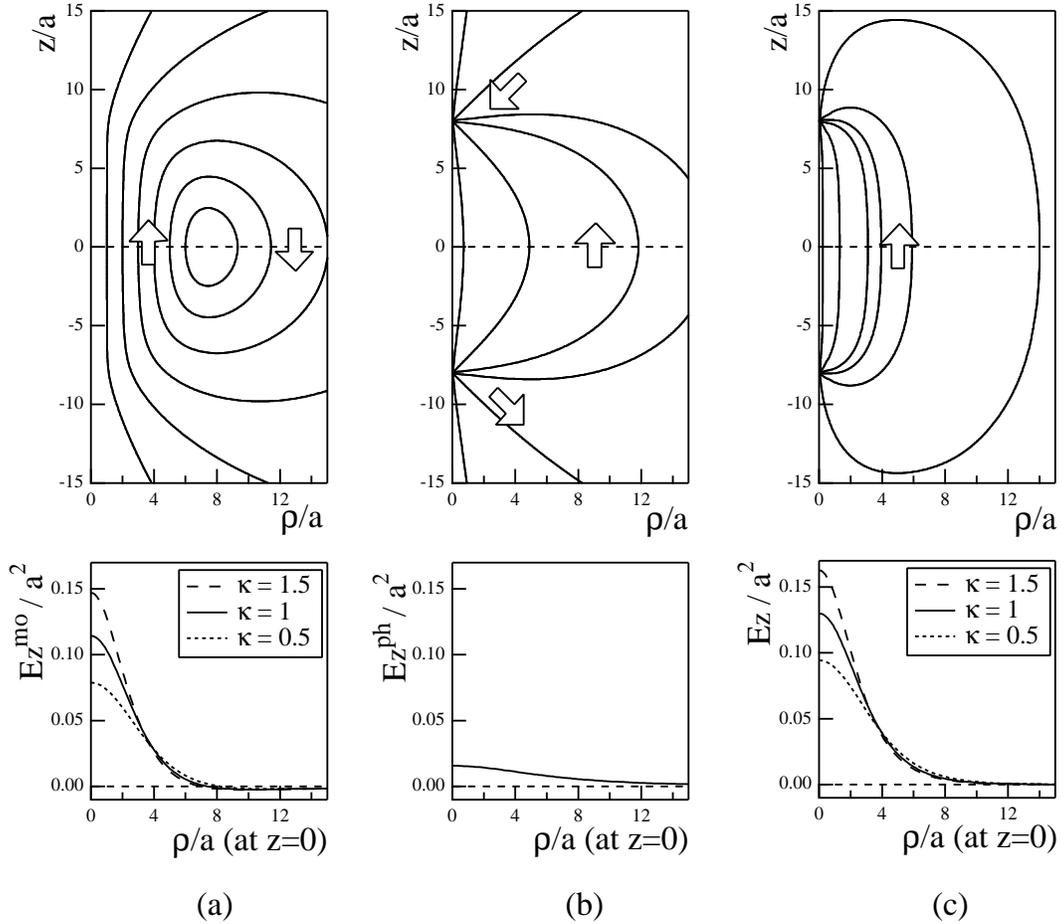}
\caption{
The flux-line pattern of the electric field (upper row) 
and the electric field strength as a function of 
the cylindrical radius (lower row):
(a) the solenoidal electric field $\bvec{E}^{mo}$ 
and (b) the Coulombic field $\bvec{E}^{ph}$ 
add up to the flux-tube profile of the full electric field (c).}
\label{fig:electric-profile}
\end{figure}

\begin{figure}[!t]
\centering
\includegraphics[width=5cm]{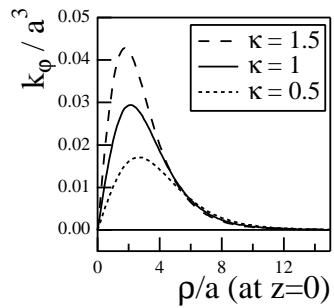}
\caption{
The monopole current strength as a function of 
the cylindrical radius.}
\label{fig:mono-profile}
\end{figure}

\par
A typical flux-tube solution,
the profile of the electric field 
and the monopole current, for the 
straight $q$-$\bar{q}$ system
is shown in Fig.~\ref{fig:dgl-profile}.
The parameters we have chosen are
$\beta_{m}=1/g^{2}=1$, $m_{B} \cdot a =\sqrt{2}g v \cdot a = 0.5$ 
and $m_{\chi}\cdot a=2\sqrt{\lambda}v\cdot a=0.5$,
taking the $q$-$\bar{q}$ separation $r=16 \;a$,
where $a$ is a certain length scale.
The Ginzburg-Landau parameter is
here $\kappa \equiv  m_{\chi}/m_{B}=1$, which means
the vacuum has superconducting properties just between
type-I and type-II vacuum.
This set of parameters is just to illustrate
the flux-tube profile as an example.
In Fig.~\ref{fig:electric-profile},
we then show the ingredient of the electric-field 
profile based on Eq.~\eqref{eq:electricprofile}.
We plot the flux-line pattern of the electric fields
along the $q$-$\bar{q}$ axis,
and the strength of each field
as a function of the cylindrical radius.
In Fig.~\ref{fig:mono-profile}, we plot
the strength of azimuthal monopole-current profile.
Here, for the plots of the electric field and 
monopole current, we have added two cases 
corresponding to
 $m_{B} \cdot a=0.5$ and $m_{\chi} \cdot a=0.25$
($\kappa=0.5$ : type I)
and $m_{B}\cdot a=0.5$ and $m_{\chi}\cdot  a=0.75$ 
($\kappa=1.5$ : type II).
Note that only the monopole-related part 
depends on $\kappa$, while the photon part does not.

\par
In Fig.~\ref{fig:electric-profile}, we find that although the 
electric field derived from the monopole part of the dual gauge 
field, $E_{z}^{mo}$, takes positive value
near the center, it turns {\em negative} beyond a certain 
radius $\rho_{c}$ (in the given case, $\rho_{c} \sim 7 a$):
this signals the appearance of a 
{\em solenoidal} electric field
which plays an important role to cancel
the Coulombic field, $E_{z}^{ph}$, 
induced by electric charges, 
at some distance from the electric Dirac string.
By this interplay the total electric field 
is squeezed from the dual superconducting vacuum, which
finally leads to a flux tube.
This is the {\em composed internal structure} of the DAH flux tube
we are referring to. The solenoidal electric field 
and monopole supercurrent are
related by the relation, $\rot{\bvec{E}}^{mo} = \bvec{k}$.
It is important to realize that 
although the shape of total electric field profile 
becomes steeper as increasing $\kappa$
due to the change of its monopole part,
the flux tube is always composed of 
the Coulombic and solenoidal electric fields.
For the infinitely separated $q$-$\bar{q}$ system,
the Coulombic contribution  disappears 
and only the solenoidal electric field remains,
where translational invariance of the flux-tube 
profile along the $q$-$\bar{q}$ axis becomes manifest.

\par
Now we come to the main point of the present section.
Through the path-integral duality transformation, 
which has formally led us to the DAH model, 
we have found the role of the photon Wilson loop 
$W_{\mathit{Ph}}[j]=\exp[i(\theta^{\mathit{ph}},j)]$
and the monopole Wilson loop 
$W_{\mathit{Mo}}[j]=\exp[i(\theta^{\mathit{mo}},j)]$
for the DAH model and its flux-tube solution; 
$W_{\mathit{Ph}}[j]$ leads to the square of the Coulombic field
strength $C^{(e)}$ after the integration over
$\theta^{ph}$ (see, Eq.~\eqref{eq:C_e_definition}), 
while $W_{\mathit{Mo}}[j]$
is translated into the interaction term between 
$\tilde{\theta}^{ph}= 2 \pi \Delta^{-1} \delta n^{(e)}$ 
and the monopole field $\chi$ 
(see, Eqs.~\eqref{eq:intersection} 
and~\eqref{eqn:currentsum}).
Namely, the photon Wilson loop provides 
the origin of the Coulombic electric field 
contribution to the DAH flux tube.
On the other hand, the monopole Wilson loop
determines the non-trivial behavior of the dual gauge field,
inducing the monopole supercurrent 
and the solenoidal electric field component
of the DAH flux tube.
The Coulombic and solenoidal electric fields 
are responsible for the Coulombic and linearly rising parts
of the inter-quark potential in the DAH model.
In the actual AP lattice gauge simulations,
it has been numerically shown that the potential 
detected by the photon and monopole 
Wilson loops have just the same 
feature~\cite{Stack:1994wm,Shiba:1994ab,Bali:1996dm}.
Now, this is naturally understood
from the relation between each Wilson loop 
and the composed internal structure of the DAH flux tube.
We then expect that the AP flux tube 
will exhibit the same composed structure as 
in the DAH flux tube, where
$W_{\mathit{Ph}}[j]$ and $W_{\mathit{Mo}}[j]$
would be respective sources.

%%%%%%%%%%%%%%%%%%%%%%%%%%%%%%
\section{Detecting the composed structure of 
Abelian-projected flux tube}
\label{sec:3}

\par
In this section, we are going to confirm 
the composed structure of the AP flux tube
by measuring the electric field and monopole current
profiles induced from the photon and monopole parts
of the Abelian Wilson loop,
based on the Monte-Carlo simulation
of SU(2) lattice gauge theory in the MAG.

\par
In order to measure the 
the flux-tube profile induced by the 
Abelian Wilson loop $W_{A}[j] =\exp [i(\theta,j)]$,
one can schematically use the following 
relation for a local operator ${\cal O}$:
\bea
\langle {\cal O} \rangle_{j} 
&=&
\frac{\int_{-\pi}^{\pi} {\cal D} \theta  
\sum_{n^{(m)}}
\; {\cal O} \; \exp \left [
- \frac{1}{2} (F,\Delta D F) + i(\theta,j) \right ] }
{\int_{-\pi}^{\pi} {\cal D}\theta  \sum_{n^{(m)} } \; 
\exp \left [ - \frac{1}{2} (F,\Delta D F)  + i(\theta,j) \right ]}
\nonumber\\*
&=&
\frac{\int_{-\pi}^{\pi} {\cal D}\theta   \sum_{n^{(m)}} 
\; {\cal O} W_{A}[j]  \; \exp \left [
- \frac{1}{2} (F,\Delta D F)  \right ] }
{\int_{-\pi}^{\pi} {\cal D}\theta
\sum_{n^{(m)}} \;  W_{A}[j] \; 
\exp \left [ - \frac{1}{2} (F,\Delta D F) \right ]}
\nonumber\\*
&=&
\frac{\langle {\cal O} W_{A}[j] \rangle_{0} }
{\langle W_{A}[j] \rangle_{0} },
\label{eq:expectation}
\eea
where $\langle \cdots \rangle_{j}$ denotes an average 
in the vacuum with an external source,
and $\langle \cdots \rangle_{0}$ an average in the vacuum 
without such source.
Thus by measurement of the expectation values of 
$\langle {\cal O} W_{A} \rangle_{0}$ and 
the Abelian Wilson loop $\langle W_{A} \rangle_{0}$,
the expectation value of a local operator associated with 
the external source, $\langle {\cal O} \rangle_{j}$, 
can be evaluated.
Below, the Abelian field strength $F$ and 
the monopole current $k$ have been chosen as
local operators ${\cal O}$.
In the actual simulation, 
since we do not know the exact form of the AP action,
we first generate non-Abelian SU(2) 
gauge configurations and then
specify the U(1) degrees of freedom
by Abelian projection after MAG fixing.

\par
Typical profiles of the electric field and monopole current
measured in this context are shown in Fig.~\ref{fig:apsu2-profile}
(some details of the simulation are given below briefly
and in Appendix~\ref{sec:simulation}).
Already at glance, the shape of the resulting profiles are 
very similar to the flux-tube profiles 
obtained within the DAH model, see Fig.~\ref{fig:dgl-profile}.

\begin{figure}[!t]
\centering
\includegraphics[width=12cm]{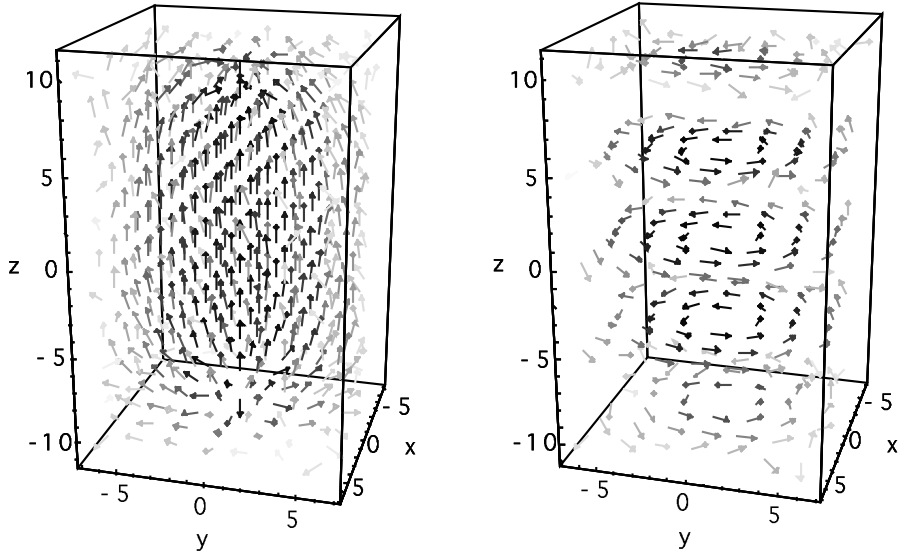}
\caption{Profiles of electric field (left) 
and monopole current (right) at
$\beta=2.5115$, with an Abelian Wilson loop
of size 16$\times$4 on a $32^{4}$ lattice.}
\label{fig:apsu2-profile}.
\end{figure}

\par
Before discussing the numerical simulation further,
it is useful to consider what happens
if we insert $W_{\mathit{Ph}}$ and $W_{\mathit{Mo}}$ 
into Eq.~\eqref{eq:expectation} instead of $W_{A}$.
Using Eq.~\eqref{eq:wilson-decomposition}
and writing the Abelian field strength 
as $F = d \theta^{ph} + 2 \pi C^{(m)} 
\equiv F_{\mathit{Ph}} +F_{\mathit{Mo}}$, 
we can expect
\bea
\langle F  \rangle_{j} 
&=&
\frac{\langle (F_{\mathit{Ph}}+F_{\mathit{Mo}}) 
W_{\mathit{Ph}}[j] W_{\mathit{Mo}}[j] \rangle_{0} }
{\langle W_{\mathit{Ph}}[j] W_{\mathit{Mo}}[j] \rangle_{0} }
\approx
\frac{\langle F_{\mathit{Ph}} W_{\mathit{Ph}}[j] \rangle_{0} }
{\langle W_{\mathit{Ph}}[j] \rangle_{0}}
+
\frac{\langle F_{\mathit{Mo}} W_{\mathit{Mo}}[j] \rangle_{0} }
{\langle W_{\mathit{Mo}}[j] \rangle_{0} }
=
\langle F_{\mathit{Ph}} \rangle_{j} 
+ \langle F_{\mathit{Mo}} \rangle_{j} \; .
\label{eq:coulomb_solenoid_separation}
\eea
Here, we have taken into account that 
in many cases lattice simulations in the MAG
have found operators
$X_{\mathit{Ph}}$ and $Y_{\mathit{Mo}}$,
defined in terms of the photon part and the monopole part
of the Abelian link variable $\theta$, respectively,
to be uncorrelated:
$\langle X_{\mathit{Ph}} Y_{\mathit{Mo}} \rangle_{0} 
\approx \langle X_{\mathit{Ph}}  \rangle_{0} 
\langle Y_{\mathit{Mo}} \rangle_{0}$
(see, {\it e.g.}, Ref.~\cite{Bali:1996dm} 
and references therein).
From the relation (\ref{eq:coulomb_solenoid_separation}),
we expect that the sum of the flux profiles 
induced by the photon and monopole Wilson loops
reproduces the total AP flux tube.

\begin{figure}[!t]
\includegraphics[height=6.5cm]{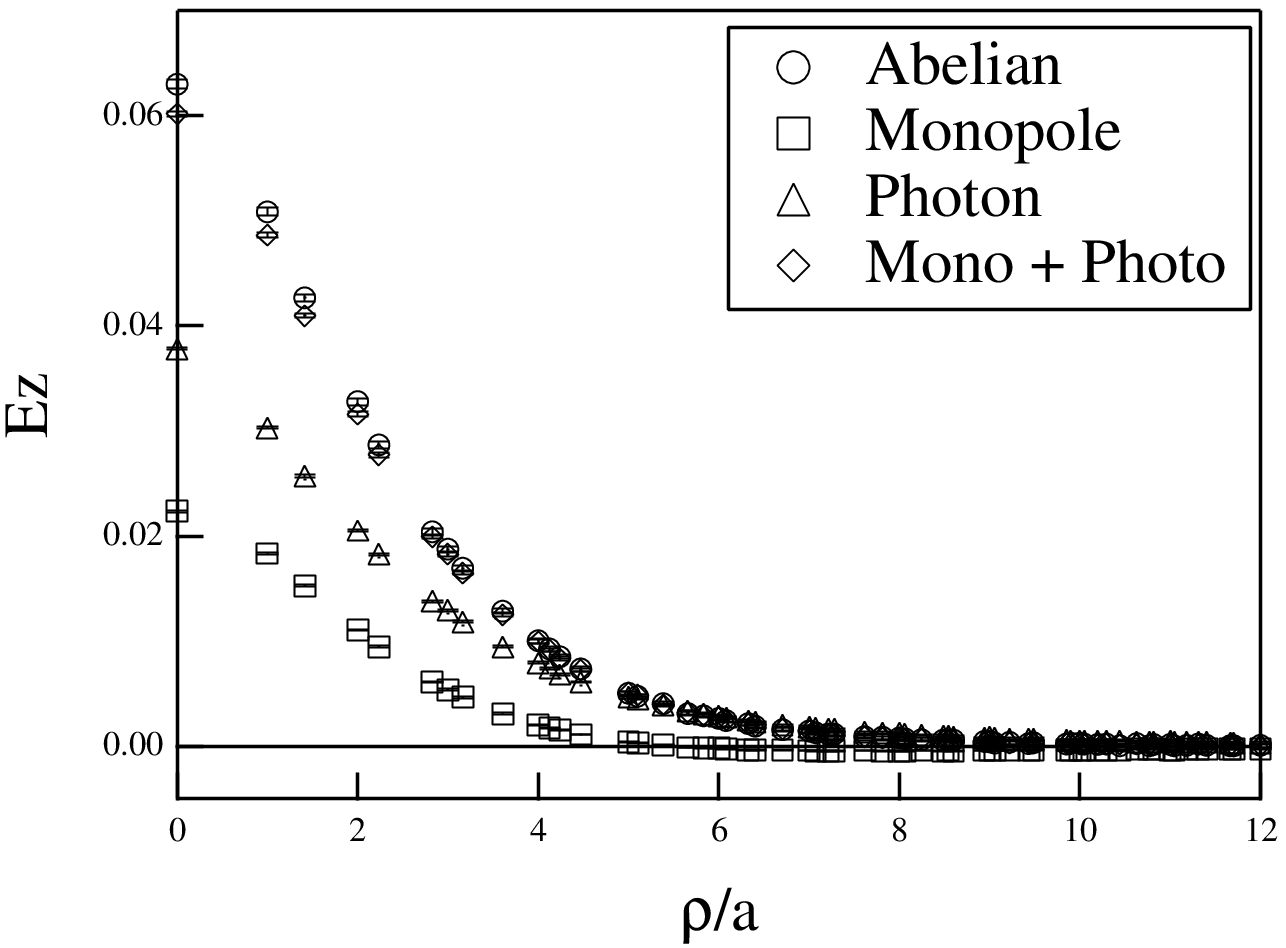}
\includegraphics[height=6.5cm]{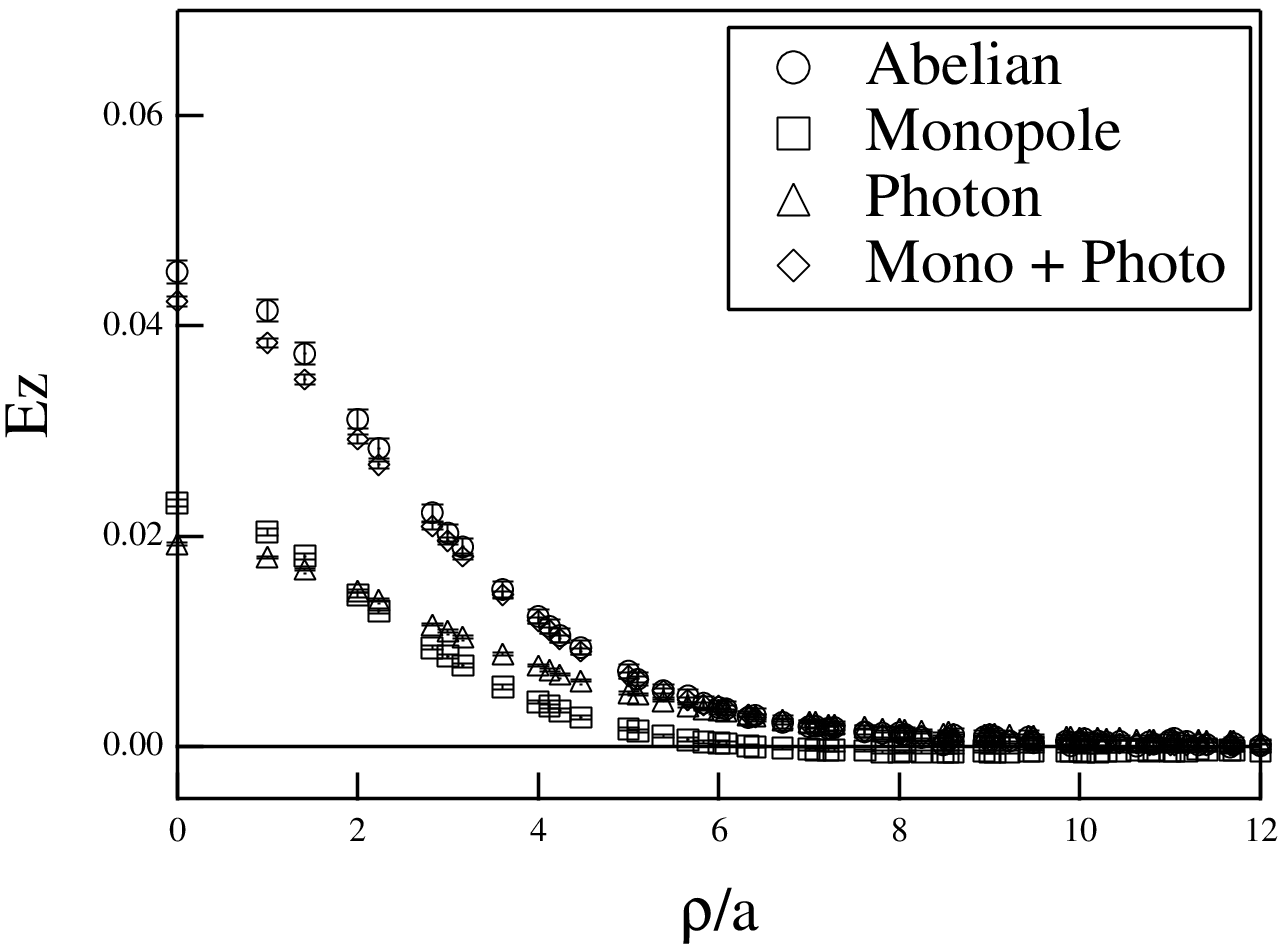}
\caption{Electric field profile from correlators with 
Abelian, photon and monopole Wilson loops
at $r=6a=0.48$ fm (left) and at $r=12a=0.97$ fm (right).}
\label{fig:electric_structure1}
\end{figure}

\begin{figure}[!t]
\includegraphics[height=6.5cm]{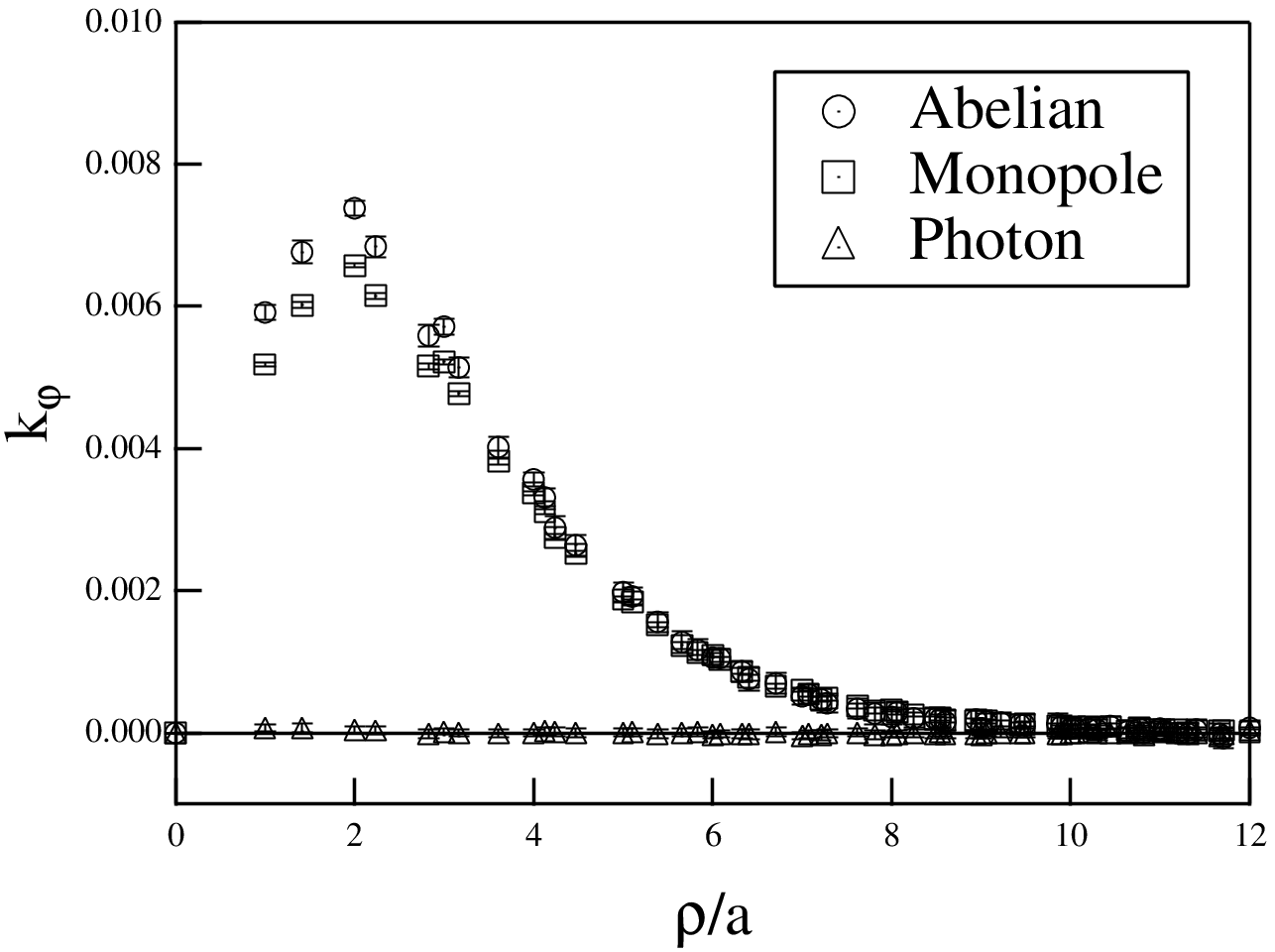}
\includegraphics[height=6.5cm]{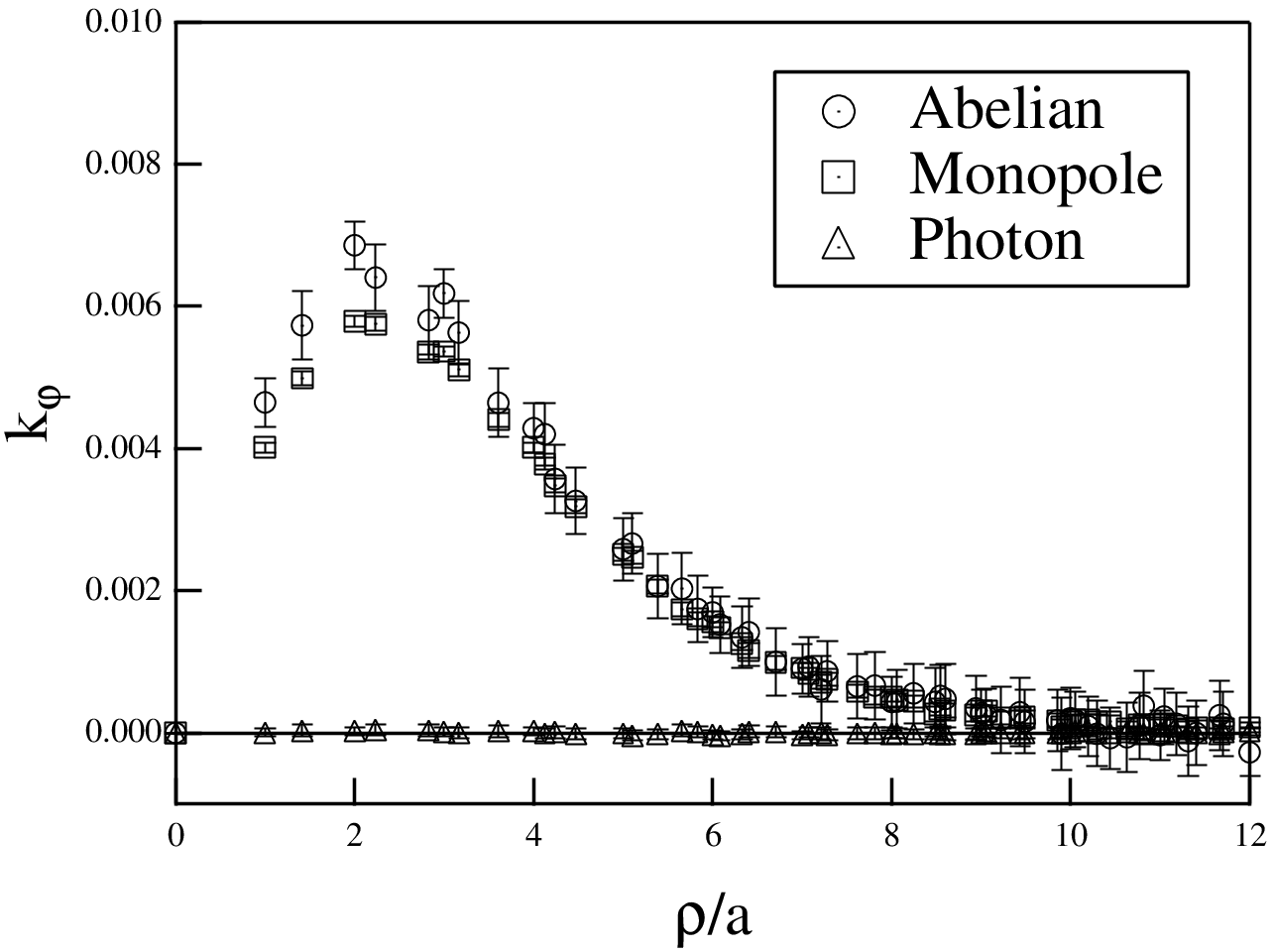}
\caption{Monopole current profile from correlators with 
Abelian, photon and monopole Wilson loops
at $r=6a=0.48$ fm (left) and at $r=12a=0.97$ fm (right).}
\label{fig:monopole_structure}
\end{figure}

\begin{figure}[!t]
\includegraphics[height=6.5cm]{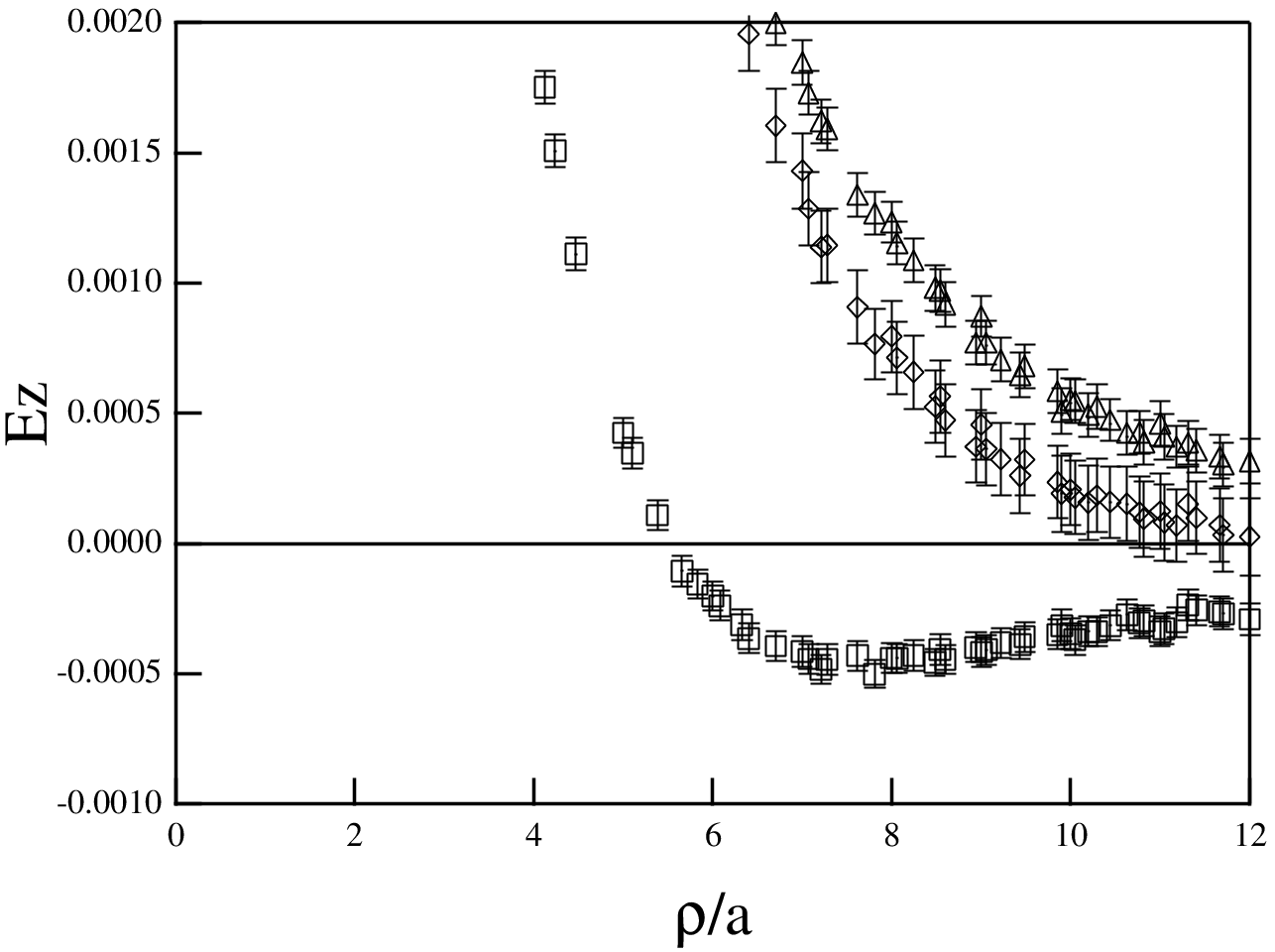}
\includegraphics[height=6.5cm]{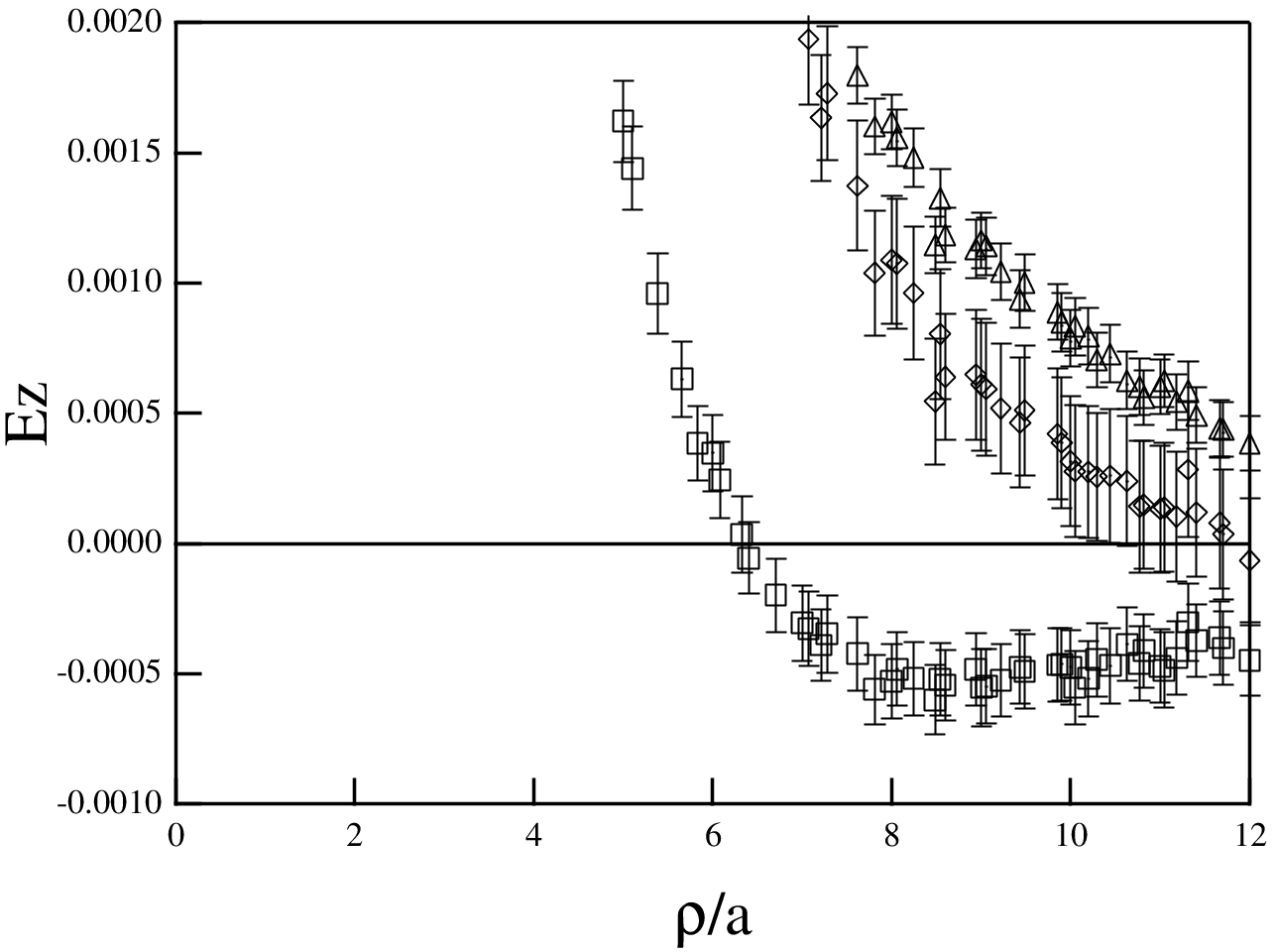}
\caption{The same plots 
as in Fig.~\ref{fig:electric_structure1}, 
for flux tubes of length $r=0.48$ fm and $r=0.97$ fm,
with the $E_{z}$ axis rescaled.  
The profile directly from the
Abelian Wilson loop is omitted.}
\label{fig:electric_structure2}
\end{figure}

\par
Second, let us consider the expectation value of the 
monopole current $k$.
Since we have the obvious relation
$k  = - d C^{(m)} =  k_{\mathit{Mo}}$,
where a photon part of the monopole current does not exist,
$k_{\mathit{Ph}} \propto  d^{2} \theta^{ph}= 0$, we will observe
\be
\langle k \rangle_{j} \approx 
\frac{\langle k_{\mathit{Mo}} W_{\mathit{Mo}}[j] 
\rangle_{0}}{\langle W_{\mathit{Mo}}[j] \rangle_{0}}
=
\langle k_{\mathit{Mo}} \rangle_{j}.
\label{eq:mono_current}
\ee
This means that the correlator  
of the monopole current only with the 
monopole Wilson loop will account for the 
full expectation value of monopole current profile and, 
at the same time, the correlator 
with the photon Wilson loop vanishes everywhere.

\par
We then show the corresponding lattice results,
the electric field profile
in Fig.~\ref{fig:electric_structure1} 
and 
the monopole current profile 
in Fig.~\ref{fig:monopole_structure},
both as a function of the cylindrical radius.
These measurements have been done at $\beta=2.5115$
on a $32^{4}$ lattice after the MAG has been fixed.
The $q$-$\bar{q}$ distances are $r=6a$ and $12a$,
and the measurements refer to the $x$-$y$ plane
at half-distance.
The lattice spacing is $a = 0.081$ fm, which  
has been determined from the non-Abelian 
string tension $\sigma_{L}$, 
$\sqrt{\sigma_{phys}} = \sqrt{\sigma_L}/a \equiv 440$ MeV. 
Physically, the $q$-$\bar{q}$ distances correspond to
0.48 fm and 0.97 fm, respectively 
(see, Appendix~\ref{sec:simulation}).
In Fig.~\ref{fig:electric_structure2} we show the 
same electric field 
profiles as in Fig.~\ref{fig:electric_structure1}, focussing
on the region where the monopole part of $E_{z}$ becomes 
negative.

\par
We find that these lattice results
concerning the behavior of the profiles 
strongly support our considerations above; 
from the photon and the monopole Wilson loops, we obtain 
the Coulombic electric field and the solenoidal electric 
field with the monopole supercurrent profile, respectively. 
We find that the sum of these two contribution 
reproduces the profile obtained from the complete Abelian 
Wilson loop (see, Eq.~\eqref{eq:coulomb_solenoid_separation}).
There is no correlation between the photon Wilson loop and 
monopole current as 
anticipated in Eq.~\eqref{eq:mono_current}.
Hence, we conclude that the AP flux tube has the same 
composed structure as the DAH flux tube.

\par
The behavior of the profiles as a function of the
$q$-$\bar{q}$ distance $r$ is also remarkable.
While the monopole Wilson loop contributions,
the solenoidal electric field and the monopole current profiles
in the midplane
are rather stable with respect to $r$,
the photon Wilson loop contribution ({\it i.e.} the Coulombic
electric field) drastically 
changes. From Fig.~\ref{fig:electric_structure1} it 
becomes obvious
that the latter determines the change of the 
full Abelian electric field profile for different $r$.
In order to see a really
translationally invariant
profile of the electric field, 
we need practically infinite $q$-$\bar{q}$ 
separation, $r \to \infty$. 
In this limit, the profile only from 
the monopole part remains.
This situation is also the same in the DAH flux tube.

%%%%%%%%%%%%%%%%%%%%%%%%%%%%%%
\section{Summary}
\label{sec:4}

\par
It has already been known 
that the profiles of the classical flux-tube 
solution in the dual Abelian Higgs 
(DAH) model and of the Abelian-projected (AP) flux tube, 
observed in lattice simulations 
in the maximally Abelian gauge (MAG), look quite similar.

\par
In this paper, in order 
to establish a more detailed correspondence 
between these two kinds of profiles,
we have studied the composed structure of both flux tubes
more carefully.
First,  by applying the path-integral duality transformation
to the Villain type compact QED
considered as the approximate action of the AP gauge theory,
we have been led to the U(1) DAH model.
Along the way, we have identified the electric and magnetic parts 
of the Abelian Wilson loop by the Hodge decomposition,
and have clarified the role of each 
contribution to the structure of the flux-tube
solution in the DAH model.
The photon and monopole Wilson loops provide
sources of the Coulombic and solenoidal electric field 
components of the DAH flux tube.

\par
Guided by this observation,
we have performed lattice simulations
of the SU(2) lattice gauge theory in the MAG
and have measured the flux profiles induced by 
the photon and the monopole Wilson loops.
We have found that the resulting profiles
show the same composed structure as 
the  DAH flux tube.

\par
The further question would be how both 
sides are related quantitatively.
One way would be to fit the profile of the 
AP flux tube by that of the DAH flux tube
and to determine the DAH parameters which 
remain unknown in the formal derivation
of the DAH model.
Here, we would like to emphasize that 
the composed structure of the AP 
flux tube found here and its relation to the DAH flux tube 
will be important for further quantitative discussions.
In fact, there is no such a work that takes into account 
the correspondence of the structures.
In addition to this, 
as we have mentioned briefly in the introduction,
a more systematic study of 
the AP flux-tube profile itself is required: 
the Gribov copy effect in the MAG, the scaling property, 
the $q$-$\bar{q}$ distance dependence, etc.
Otherwise, one cannnot trust the robustness and 
physical relevance of the resulting DAH parameters.
A part of such a quantitative analysis is 
reported in Lattice 2002~\cite{Koma:2002uq}
and the detailed report 
will be presented in Ref.~\cite{koma-next}.

\par 
In closing, we note that although we have concentrated 
here on SU(2) gauge theory, the ideas
discussed in the present paper
can be extended to arbitrary AP-SU($N$) gauge theory in the 
MAG~\cite{Bornyakov:2001nd,Koma:2002cv,Ichie:2002mi}.

%%%%%%%%%%%%%%%%%%%%%%%%%%%%%%%%%%%%%%%%%%%
\begin{acknowledgments}
We are grateful to V. Bornyakov, H. Ichie, G. Bali, 
R.~W.~Haymaker, D. Ebert, and M.~N.~Chernodub
for useful discussions.

Y.~K. was partially supported  by
the Ministry of Education, Science, Sports and Culture,
Japan (Monbu-Kagaku-sho), Grant-in-Aid for Encouragement 
of Young Scientists (B), 14740161, 2002.

E.-M.~I. acknowledges gratefully the support by Monbu-Kagaku-sho 
which allowed him to work within the COE program 
at Research Center for Nuclear Physics 
(RCNP), Osaka University, where this work has begun.
He expresses his personal thanks to H.~Toki for the hospitality.
E.-M.~I. is presently supported by DFG through the DFG-Forschergruppe
'Lattice Hadron Phenomenology' (FOR 465).

T.~S. is partially supported by JSPS Grant-in-Aid for Scientific 
Research on Priority Areas No.13135210 and (B) No.15340073.

M.~I.~P. is partially supported by grants RFBR 02-02-17308, RFBR
01-02-17456, DFG-RFBR 436 RUS 113/739/0,
INTAS-00-00111 and CRDF award RPI-2364-MO-02.

The computation was done on the Vector-Parallel Supercomputer 
NEC SX-5 at the RCNP, Osaka University, Japan.

\end{acknowledgments}

%%%%%%%%%%%%%%%%%%%%%%%%%%%%%%%%%%%%%%%%%%%
\appendix
\section{Classical flux-tube solution in the DAH model}
\label{sec:DAHfluxtube}

In this appendix, we present concrete form of 
the field equations of the DAH model, 
Eqs.~(\ref{eq:feq-1-m}) and (\ref{eq:feq-2-m}),
for a straight $q$-$\bar{q}$ system.
We use here the continuum notations;
let $B^{mo}_{\mu}$ be the continuum form of 
the regular dual gauge field
denoted $\tilde{\theta}_{\mu}^{mo}$ on the lattice, 
$B^{ph}_{\mu}$ that of the singular dual gauge field
$\tilde{\theta}_{\mu}^{ph}$.

\par
We put the quark and the antiquark at  
$\bvec{x}_{1} = (-r/2) \bvec{e}_{z}$
and $\bvec{x}_{2} = (+r/2) \bvec{e}_{z}$.
The corresponding electric current is written as
$j_{\mu}(x)=\delta_{\mu 0}
\left( \delta(\bvec{x}-\bvec{x}_1)
-\delta(\bvec{x}-\bvec{x}_2) \right)$. 
Since this system has cylindrical geometry, 
the fields can be parametrized in terms of 
cylindrical coordinates $(\rho,\varphi,z)$ as
\bea
\phi &=& \phi(\rho,z)\; ,\\*
\bvec{B}^{mo} &=&
B^{mo}(\rho,z)\bvec{e}_{\varphi} 
\equiv  
\frac{\hat{B}^{mo}(\rho,z)}{\rho}\bvec{e}_{\varphi}\; ,\\*
\bvec{B}^{ph} &=& 
-\frac{n}{2 \rho}
\left ( 
 \frac{z+r/2}{\sqrt{\rho^2 +(z+r/2)^2}}
-\frac{z-r/2}{\sqrt{\rho^2 +(z-r/2)^2}}
\right ) \bvec{e}_{\varphi}\;  .
\label{eqn:b-sing-fin}
\eea
The factor $n$ in $\bvec{B}^{ph}$ is
the winding number of the flux tube (an integer value), 
which is determined by the representation of the electric 
charges~\cite{Koma:2001ut,Koma:2001is,Koma:2002cv}.
The fundamental representation corresponds to $n=1$.

\par
The field equations (\ref{eq:feq-1-m}) and (\ref{eq:feq-2-m}) 
are then reduced to:
\bea
&&
\beta_{m}
\left (
\frac{\partial^2 \hat{B}^{mo}}{\partial \rho^2} 
-\frac{1}{\rho}\frac{\partial \hat{B}^{mo}}{\partial \rho}
+ \frac{\partial^2 \hat{B}^{mo}}{\partial z^2} 
\right )
\nonumber\\*
&&
-2 \left (\hat{B}^{mo}  - \frac{n}{2} 
\left (\frac{z+r/2}{\sqrt{\rho^2 + \left ( z+r/2 \right )^2 }}
-\frac{z-r/2}{\sqrt{\rho^2 + \left ( z-r/2 \right )^2 }}
\right ) \right ) \phi^2 = 0,\\
&&
\frac{\partial^2 \phi}{\partial \rho^2} 
+ \frac{1}{\rho}\frac{\partial \phi}{\partial \rho}
+ \frac{\partial^2 \phi}{\partial z^2}\nonumber\\*
&&
-
\left ( \frac{\hat{B}^{mo} - \frac{n}{2} 
\left (\frac{z+r/2}{\sqrt{\rho^2 + \left ( z+r/2 \right )^2 }}
-\frac{z-r/2}{\sqrt{\rho^2 + \left ( z-r/2 \right )^2 }}\right ) 
}{\rho} \right )^2 \phi
- 2 \lambda \phi (\phi^2 - v^2) = 0.
\eea
The boundary conditions are specified
so as to make the energy of the system finite as
\bea
&&
\hat{B}^{mo} = 0 
\quad {\rm as} \quad \rho \to 0, 
\quad {\rm and} \quad \phi = 0
\quad {\rm as} \quad \rho \to 0 \quad {\rm for} \quad -r \leq z \leq r, 
\nonumber\\*
&&
\hat{B}^{mo} =  
\frac{n}{2}
\left ( \frac{z+r/2}{\sqrt{\rho^2 + \left ( z+r/2 \right )^2 }}
-\frac{z-r/2}{\sqrt{\rho^2 + \left ( z-r/2 \right )^2 }}\right )
\quad {\rm and} \quad  \phi = v
\quad {\rm as} \quad  \rho, \; z  \to \infty \; .
\label{eqn:dah-boundary-condition}
\eea
After getting the numerical solution of 
the field equations for $\hat{B}^{mo}$ and 
$\phi$, the profiles of the electric field are computed
as Eq.~\eqref{eq:electricprofile}, where
\bea
\bvec{E}^{mo} 
&=&
-\frac{1}{\rho} \frac{\partial \hat{B}^{mo}}{\partial z}
\bvec{e}_{\rho}
+\frac{1}{\rho}\frac{\partial \hat{B}^{mo}}{\partial \rho}
\bvec{e}_{z},
\\*
\bvec{E}^{ph}
&=&
\frac{n}{2}\left (
\frac{\rho}{(\rho^{2} +(z +r/2)^{2})^{3/2}}
-
\frac{\rho}{(\rho^{2} +(z -r/2)^{2})^{3/2}}
\right )\bvec{e}_{\rho}
\nonumber\\*
&&
+
\frac{n}{2}\left (
\frac{z+r/2}{(\rho^{2} +(z +r/2)^{2})^{3/2}}
-
\frac{z-r/2}{(\rho^{2} +(z -r/2)^{2})^{3/2}}
\right ) \bvec{e}_{z}.
\eea
The profile of the monopole current~\eqref{eq:magneticprofile}  is 
given by 
\bea
\bvec{k}
=
2 \left (\hat{B}^{mo}  - \frac{n}{2} 
\left (\frac{z+r/2}{\sqrt{\rho^2 + \left ( z+r/2 \right )^2 }}
-\frac{z-r/2}{\sqrt{\rho^2 + \left ( z-r/2 \right )^2 }}
\right ) \right ) \phi^2 \; \bvec{e}_{\varphi} \;.
\eea

%%%%%%%%%%%%%%%%%%%%%%%%%%%%
\section{lattice simulation detail}
\label{sec:simulation}

For the SU(2) link variables, $U_{\mu}(s)$, 
generated by Monte-Carlo method with Wilson gauge action,
we adopt the maximally Abelian gauge (MAG)
fixing, which is achieved by maximizing the functional
\be
R[U^V] = \sum_{s,\mu} {\rm tr} \left \{  \tau_3 U^V_{\mu}(s) 
\tau_3 U^{V~\dagger}_{\mu}(s) \right \} \; .
\label{eq:gauge_functional}
\ee
After the MAG fixing, Abelian projection is performed; 
the SU(2) link variables $U_{\mu}^{V}(s)=U_{\mu}^{MA}(s)$
are factorized into a diagonal (Abelian) link 
variable $u_{\mu}(s) \in {\rm U(1)}_3$ 
and the off-diagonal (charged matter field) parts 
$c_{\mu}(s)$, $c_{\mu}^*(s) \in {\rm SU(2)}/{\rm U(1)}_3$ 
as follows
\bea
U_{\mu}^{\mathit{MA}}(s)=
\left (
\begin{array}{cc}
\sqrt{1-|c_{\mu}(s)|^2} & -c_{\mu}^*(s)\\
c_{\mu}^*(s) & \sqrt{1-|c_{\mu}(s)|^2}
\end{array}
\right )
\left (
\begin{array}{cc}
u_{\mu}(s) & 0 \\
0 & u_{\mu}^*(s) 
\end{array}
\right ) \; ,
\label{eqn:lat-Cartan-decompose}
\eea
where the Abelian link variables  $u_{\mu}(s)$  are then 
explicitly written as
\bea
u_{\mu}(s) =e^{i\theta_{\mu}(s)}
\quad  (\theta_{\mu}(s) \in [-\pi, \pi )  )\; .
\label{eq:Abelian_links}
\eea
The Abelian plaquette variables is then constructed as
\bea
\theta_{\mu\nu}(s)
&\equiv& 
\theta_{\mu}(s) +\theta_{\nu}(s+\hat{\mu})
- \theta_{\mu}(s+\hat{\nu})-\theta_{\nu}(s)
 \quad \in [-4\pi, 4 \pi) \; ,
\label{eq:Abelian_plaquette}
\eea
which is decomposed into a regular part
$\bar{\theta}_{\mu\nu}(s) \in [-\pi,\pi)$
and a singular (magnetic Dirac string) 
part $n_{\mu\nu}^{(m)}(s) = 0,\pm 1,\pm 2$  as follows
\bea
\theta_{\mu\nu}(s)
&\equiv& 
\bar{\theta}_{\mu\nu}(s) +2\pi n_{\mu\nu}^{(m)}(s) \; .
\label{eq:Abelian_plaquette_decomposition}
\eea
The Abelian field strength is defined by 
$\bar{\theta}_{\mu\nu}(s) 
=\theta_{\mu\nu}(s) -2 \pi n_{\mu\nu}^{(m)}(s)$.
Following DeGrand and Touissaint~\cite{DeGrand:1980eq},
magnetic monopoles are extracted as the string boundaries
\bea
k_{\mu}(s_{d}) 
=
- \frac{1}{2} \varepsilon_{\mu\nu\rho\sigma}
\partial_{\nu} n_{\rho\sigma}^{(m)} (s+\hat{\mu})
\qquad (  \varepsilon_{1234}=1 ) \; ,
\label{eq:magnetic_current_definition}
\eea
where $|k_{\mu}(s_{d})| \leq 2$ and 
$s_{d} \equiv  s+ (\hat{1} +  \hat{2} + 
\hat{3} + \hat{4})/2$ denotes the dual site.

\par
For measuring the correlation function, 
we have used the following local operators: 
 an electric field operator
\be
{\cal O}(s) = i \bar{\theta}_{i4} (s) 
= i (\theta_{i4}(s) -2 \pi n_{i4}^{(m)}(s)), 
\label{eq:electric_field}
\ee
and a monopole current operator
\bea
{\cal O}(s_{d}) = 2 \pi i k_{i}(s_{d}) 
\label{eq:monopole_current}
\eea
The Abelian Wilson loop is constructed as
\be
W_{A}[j]=\prod_{l \in j}^{} u_{\mu}(s)= e^{i \sum_{l \in j}^{} 
\theta_{\mu}(s)}.
\ee
Similarly, the photon and the monopole Wilson loop
are constructed from the photon and monopole parts 
of Abelian link variables, $\theta^{ph}$ and $\theta^{mo}$,
respectively, where
\bea
\theta_{\mu}(s) 
=  \Delta^{-1}\partial_{\nu} 
(\bar{\theta}_{\mu\nu}(s)+2\pi  n_{\mu\nu}^{(m)}(s))
=\theta_{\mu}^{ph}(s) + \theta_{\mu}^{mo}(s).
\eea
In this decomposition, it is necessary to adopt
the Abelian Landau gauge  which is 
characterized by $\partial_{\mu} \theta_{\mu}(s)=0$.
Note, however, that the Wilson loops constructed from each
link variable are Abelian gauge invariant.

\par
In this simulation, in  order to see the profiles which
belong to the ground state of a flux tube,
we have adopted a smearing technique for 
spacelike Abelian link variables.
Then we have constructed the {\em smeared} Abelian
Wilson loop~\cite{Bali:1996dm}.
Considering the fourth direction as the Euclidean 
time direction, we have performed $N_{s}$ times 
the following step in a smearing procedure 
applied only to the {\it spatial} 
Abelian links ($i,j = 1,2,3$), 
\be
 \alpha e^{i \theta_{i}(s)}
+ \sum_{j \ne i}
e^{i (\theta_{j}(s) +\theta_{i}(s+ \hat{j})
- \theta_{j}(s+\hat{i}) )}
\to e^{i \theta_{i}(s)} \; ,
\label{eq:smearing_step}
\ee
where $\alpha$ is an appropriate smearing parameter.
The same procedure was also applied to the spatial parts
of the photon and the monopole link variables before
constructing each type of Wilson loop.

\par
The numerical simulations which are presented in this paper
have been done at $\beta=$ 2.5115.
The lattice volume was $32^4$.
We have used 100 configurations for measurements.
We have produced them after 3000 thermalization sweeps,
separated by 500 Monte Carlo updates. 
They have been stored for performing MAG fixing.
This has been repeated $N_g$ times, starting each time 
from a different random gauge copy of the configuration,
in order to explore an increasing number of Gribov copies.
The copy reaching the maximal value of the
gauge functional (\ref{eq:gauge_functional}) has been
selected for measuring the profiles and kept for further 
increasing of $N_g$.
Finally we have chosen $N_{g}=20$.
For the MAG fixing itself, we have 
used the simulated annealing algorithm~\cite{Bali:1996dm},
followed by a final steepest descent relaxation. 
The size of the Wilson loops mainly studied  
(for Fig.~\ref{fig:electric_structure1},~\ref{fig:monopole_structure}
and \ref{fig:electric_structure2}) are
$R\times T$ = $6 \times 6$ and $12 \times 6$ in 
units of lattice spacing $a$.
We have measured the profiles 
in the $x$-$y$ plane orthogonal to the
Wilson loop in its midpoint.
The Abelian smearing parameters 
have been found by optimization as
$N_{s}=8$ and $\alpha = 2.0$. 
With this choice,
the profiles induced by the Abelian Wilson loop with 
timelike extensions $T=8$ and $T=6$
agree within errors.

\par
The physical scale (the lattice spacing $a(\beta=2.5115)$) 
has been determined from the 
non-Abelian string
tension $\sigma_L$ by fixing $\sqrt{\sigma_{phys}}
= \sqrt{\sigma_L}/a \equiv 440$ MeV.
The non-Abelian string tension has been 
re-evaluated by measuring 
expectation values of non-Abelian Wilson loops with an optimized 
non-Abelian smearing. 
The potential has been fitted to match the form 
$V(R) = C - A/R + \sigma_L\;R$.
The resulting string tension is $\sigma_{L}=0.0323(4)$
at $\beta=2.5115$, 
such that the corresponding lattice
spacing in physical units is $a(\beta)=0.0806(5)$ fm.

%% \bibliographystyle{h-physrev3}
%% \bibliography{koma-paper}

\end{document}